\documentstyle[twocolumn,aps]{revtex}
\begin{document}
\draft
\title{Quantum correlations are not local elements of reality\thanks{
Phys. Rev. A {\bf 59}, 113-115 (1999).}}
\author{Ad\'{a}n Cabello\thanks{Electronic address: 
fite1z1@sis.ucm.es}}
\address{Departamento de F\'{\i}sica Aplicada,
Universidad de Sevilla, 41012 Sevilla, Spain}
\date{\today}
\maketitle
\begin{abstract}
I show a situation of multiparticle entanglement 
which cannot be explained in the framework of 
an interpretation of quantum mechanics recently proposed elsewhere.
This interpretation is based on the assumption that correlations 
between subsystems of an individual isolated composed quantum 
system are real objective local properties of that system.
\end{abstract}

\narrowtext

Entanglement is not ``{\em one} but rather {\em the} 
characteristic trait of quantum 
mechanics, the one that enforces its entire departure 
from classical lines of 
thought'' \cite{Schrodinger35}. Entanglement between 
two systems is central in the argument
of incompleteness of quantum mechanics proposed by 
Einstein, Podolsky, and Rosen (EPR) \cite{EPR35} 
and it is also essential in Bell's proof that 
EPR's program to ``complete'' quantum mechanics leads 
to predictions in contradiction 
with those of quantum mechanics \cite{Bell64}.
More recently, 
the study of entanglement in systems of three or more particles has 
opened new chapters on both the fundamental and applied sides of 
quantum mechanics. Multiparticle entanglement has provided the first 
proof of Bell's theorem without inequalities \cite{GHZ89}, and it is 
ubiquitous in almost all recent developments on quantum communication 
and quantum computation.

This paper's intent is to show some difficulties
which arise when one tries to explain some situations of multiparticle 
entanglement in the framework of a recent interpretation of quantum 
mechanics proposed by Mermin in a series of 
papers \cite{Mermin96,Mermin98,Mermin98b}. 
This interpretation is based on three assumptions: 
\begin{itemize}
\item[(a)] Density matrices describe isolated 
individual systems --- not just ensembles. 
Density matrices fully describe all the internal 
correlations of an isolated individual 
system \cite{Mermina}.
\item[(b)] All correlations between subsystems of 
an isolated composed system are 
real objective internal 
properties of such subsystems \cite{Merminc}.
\item[(c)] Real objective internal properties 
of an isolated system ``cannot change in
immediate response to what is done to a far-away system 
that may be correlated but does not interact 
with the first'' \cite{Merminf}.
\end{itemize}

I will show a physical situation in which these three 
assumptions cannot be reconciled in a consistent way. 
To make this conflict clear I will assume (a) and (b), and 
then show that assumption (c) cannot be correct.

Consider the following situation. We have two sources, 
each of which emits a 
{\em single} pair of spin-$\frac{1}{2}$ particles 
in the singlet state. 
The two particles emitted by the first 
source will be labeled 1 and 2, and 
the two particles emitted by the second source will 
be labeled 3 and 4.
The initial state of the four particles is then 
\begin{eqnarray}
\left| \Psi \right\rangle _{1234}=
{1 \over 2}\left( {\left| + \right\rangle _1\otimes 
\left| - \right\rangle _2
-\left| - \right\rangle _1\otimes \left| + 
\right\rangle _2} \right) \nonumber \\ 
\otimes \left( {\left| + \right\rangle _3\otimes \left| - 
\right\rangle _4-\left| - 
\right\rangle _3\otimes
\left| + \right\rangle _4} \right).
\label{doublesinglet}
\end{eqnarray}
Let me now consider two alternative experiments:

{\em Experiment 1:} On particles 2 and 3, we perform a
measurement of component $z$ of
the spin of each particle. This measurement
projects \cite{projects} the combined state of a single 
pair of particles 2 and 3 
onto one of the following four factorizable pure states:
\begin{equation}
\left| + \right\rangle _2\otimes \left| + \right\rangle _3,
\;\;\, 
\left| + \right\rangle _2\otimes \left| - \right\rangle _3,
\;\;\, 
\left| - \right\rangle _2\otimes \left| + \right\rangle _3,
\;\;\, 
\left| - \right\rangle _2\otimes \left| - \right\rangle _3. 
\label{states23}
\end{equation}
This measurement on particles 2 and 3 also projects the 
combined state of the 
corresponding single pair of
particles 1 and 4 onto, respectively, one of the 
following factorizable pure states: 
\begin{equation}
\left| - \right\rangle _1\otimes \left| - \right\rangle _4,
\;\;\, 
\left| - \right\rangle _1\otimes \left| + \right\rangle _4,
\;\;\, 
\left| + \right\rangle _1\otimes \left| - \right\rangle _4,
\;\;\, 
\left| + \right\rangle _1\otimes \left| + \right\rangle _4. 
\label{states14}
\end{equation}
There is a one-to-one correspondence between the four 
states (\ref{states23}) and the four states (\ref{states14}): 
if the measurement on 
particles 2 and 3 projects their state onto $\left| + 
\right\rangle _2\otimes \left| + 
\right\rangle _3$, then the state of particles 
1 and 4 is projected onto 
$\left| - \right\rangle _1\otimes \left| - 
\right\rangle _4$, etc.

{\em Experiment 2:} Instead of a spin measurement on 
each particle 2 and 3, we 
perform a measurement of the Bell operator \cite{BMR92} 
on particles 2 and 3. This 
measurement projects the combined state of a single pair 
of particles 2 and 3 onto one of 
the four Bell states \cite{BMR92}
\begin{equation}
\left| {\Psi ^\pm } \right\rangle _{23}=
{1 \over {\sqrt 2}}
\left( {\left| + \right\rangle _2\otimes \left| - 
\right\rangle _3\pm 
\left| - \right\rangle _2\otimes \left| + 
\right\rangle _3} \right),
\label{Bellpsi}
\end{equation}
\begin{equation}
\left| {\Phi ^\pm } \right\rangle _{23}=
{1 \over {\sqrt 2}}
\left( {\left| + \right\rangle _2\otimes \left| + 
\right\rangle _3\pm 
\left| - \right\rangle _2\otimes \left| - 
\right\rangle _3} \right),
\label{Bellphi}
\end{equation}
which form a complete basis for the combined system 
of particles 2 and 3. This 
measurement on particles 2 and 3 also
projects the combined state of the corresponding single 
pair of particles 1 and 4 onto, respectively,
one of the Bell states:
\begin{equation}
\left| {\Psi ^+ } \right\rangle _{14},
\quad
\left| {\Psi ^- } \right\rangle _{14},
\quad 
\left| {\Phi ^+ } \right\rangle _{14},
\quad 
\left| {\Phi ^- } \right\rangle _{14}. 
\label{Bellstates14}
\end{equation}
Indeed, the measurement projects the state of 
particles 2 and 3 and the state of 
particles 1 and 4 onto the same Bell state: 
if the state of particles 2 and 3 is projected onto
$\left| {\Psi ^+ } \right\rangle _{23}$, then
the state of particles 1 and 4 is projected onto 
$\left| {\Psi ^+ } \right\rangle _{14}$, etc. 
Experiments of this second kind have been previously 
considered in the context of 
``entanglement swapping'' \cite{BBCJPW93,ZZHE93,BVK98,PBWZ98}.

Before any of the alternative experiments, particles
1, 2, 3, and 4 were {\em completely isolated} 
from the rest of the universe; 
i.e., they form a system that has no external interactions or 
correlations \cite{Mermind}. 
Since, before any experiment, the state of particles 1 and 2 is 
factorizable from the state of 
particles 3 and 4, particles 1 and 2
(3 and 4) form a completely isolated subsystem of 
the system of particles 1, 2, 3, and 4. 
Before any of the alternative experiments there is 
no (nontrivial) correlations 
between any of the particles 
1 and 2 and any of the particles 3 and 4.
 
On the other hand, {\em before} any of the experiments, 
particles 2 and 3 form a 
{\em dynamically isolated} subsystem;
i.e., they have no external 
interactions \cite{Mermind}. {\em After} any of the experiments, 
particles 2 and 3 do not form a dynamically isolated system 
since they have interacted with the measuring apparatus.
If particles 1 and 4 are spacelike separated from 
the experiment performed on particles 2 and 3, then 
particles 1 and 4 cannot interact with the measuring apparatus. 
Therefore, particles 1 and 4 form a 
dynamically isolated system {\em before and after} any
of the experiments.

Mermin's interpretation assumes {\em physical locality}, defined as 
``[t]he fact that the internal correlations of a dynamically isolated 
system do not depend on any interactions experienced by other 
systems external to it'' \cite{Mermind}. However,
while any of the four possible states of 
particles 1 and 4 after an experiment of the first type given 
in Eq. (\ref{states14}) are {\em factorizable}, any of the four 
possible states after an experiment of 
the second type given in Eq. (\ref{Bellstates14}) 
are {\em maximally entangled}.
This means that while after an experiment of the first
kind (regardless of the result), 
particles 1 and 4 have their spins correlated only in 
the $z$ direction, after an experiment
of the second kind (irrespective of the result), 
particles 1 and 4 are highly correlated: 
every component of spin of particle 1 is 
correlated with other component of spin of 
particle 4, and {\em vice versa}. 
Therefore, the internal correlations between 
particles 1 and 4 are completely 
different depending on the interaction 
between particles 2 and 3 and an external 
system. Accepting assumptions (a) and (b) means, 
in this example, the violation of physical locality
as defined by Mermin.
By this violation of physical locality 
I do not mean that the internal correlations 
between particles 1 and 4 ``change'' after a spacelike separated 
experiment (this does not happen in the sense that no new internal 
correlations are ``created'' that were not 
``present'' in the reduced density 
matrix for the system 1 and 4 before
any interaction), but that the type of 
internal correlations (and therefore, 
according to Mermin, {\em the reality}) of an individual 
isolated system can be chosen at distance.

The roles of assumptions (b) and (c) in my 
argument of ``nonlocality'' are clear: (b) is a 
definition of what is
``real'' and (c) is the corresponding condition of 
locality. Allow me to emphasize the role of 
assumption (a) for this argument. Assumption (a) says that 
a density matrix describes an
{\em individual} system --- not just an ensemble.
This is crucial since there is a large difference between the mixed 
density matrix
which describes an {\em ensemble} of pairs of particles 1 and 4, and 
the pure states 
(factorizable or maximally entangled)
which describe a {\em single} pair of particles 1 and 4. 
The mixed density matrix does not change after any set of measurements 
on particles 2 and 3 (otherwise this would mean an instantaneous 
transmission of information), 
but the pure states that
describe {\em individual} pairs of particles 1 and 4
are different, depending on which experiment 
is performed on particles 2 and 3.

In its present form the interpretation proposed by Mermin is inconsistent. 
However, some parts of his proposal could be preserved in a further 
developed interpretation. No doubt exists on the fact that quantum 
correlations are fundamental, but fundamental does not necessarily 
mean {\em real}, or at least it does not mean real if this also means 
{\em local}. 
A consistent interpretation could be developed by keeping correlations 
as fundamental but avoiding to say that they are local properties.

In fact, the previous example does not cause any conceptual problem
if one accepts the Copenhagen interpretation of quantum mechanics, 
according to which 
information about a quantum system is a more basic feature than 
any ``real'' properties 
these systems might have \cite{Zeilinger98}, even if these ``real''
 properties are limited to 
internal correlations between subsystems. Quantum mechanics is 
not a theory about 
reality \cite{Peres93}, it is only a tool for predicting 
probabilities for the various possible outcomes
of an experiment on a physical system, once we specify the 
procedure for the preparation of that system. The wave function 
(or the density matrix) only represents, in Bohr's words,
``a purely symbolic procedure, 
the unambiguous physical interpretation of which in the last 
resort requires a reference to a 
complete experimental arrangement'' \cite{Bohr58}. 

Mermin's proposal can be seen as
an attempt to ``complete'' the Copenhagen interpretation.
In Mermin's interpretation, internal correlations between two 
parts of a more than two-part 
system play a similar role to the one that 
correlated observables of a two-part system played for EPR's 
attempt to ``complete'' quantum mechanics \cite{EPR35}. The obstacle
is again the same: if one insists that these elements of reality must
be {\em local}, one can find physical situations in 
which all assumptions cannot be consistently reconciled. \\

The author thanks 
Gonzalo Garc\'{\i}a de Polavieja for fruitful discussions, and 
Christopher Fuchs, David Mermin, and Asher Peres for useful feedback. 
This work was partially supported by the Universidad de Sevilla.


\begin{thebibliography}{99}
\bibitem{projects} I would just like the expression 
``this measurement projects'' to mean that the available information 
(i.e., all the information allowed in principle by nature) about 
a physical system can be updated as a result of the interaction
between this system and the measuring apparatus.
\end{thebibliography}
\end{document}